\documentstyle [11pt]{article}
\raggedright
\begin{document}     

\parindent=0.3in
\baselineskip=0.0in

\begin{center}
{\bf Dependence of Angle Resolved Photoemission Spectra of High
Temperature Superconductors on the Spin Fluctuation Susceptibility}
\end{center}

\vspace{0.3in}
\begin{center}
L. Coffey, D. Lacy, K. Kouznetsov(*), A. Erner(+)
\end{center}

\begin{center}
Department of Physics,\\
Illinois Institute of Technology,\\
Chicago, IL 60616.
\end{center}

\vspace{2.0in}

(*) Present Address: Dept. of Physics, University of California, Berkeley,
CA. \\
(+) Present Address: Dept. of Physics, Northwestern University, Evanston,
IL.

\newpage

\begin{abstract}
Angle Resolved Photoemission Spectroscopy (ARPES) measurements on high
temperature superconductors, such as BiSrCaCuO, show three main components.
These are a quasiparticle spectral peak that develops below the
superconducting transition temperature $T_{c}$, an accompanying broad
background of secondary electrons and a dip feature beside the main
quasiparticle peak. The broad background may originate from inelastic
processes in which the photoelectron emits and absorbs spin fluctuations.
Calculations of the quasiparticle spectral weight are presented
incorporating
these spin fluctuation mediated inelastic processes in which the development
of the superconducting gap $\Delta$ has been incorporated into the magnetic susceptibility
$\chi(\vec q, E)$. A dip feature develops below $T_{c}$ in the quasiparticle
spectral weight due to the shifting of spin fluctuation spectral weight,
Im$\chi(\vec q, E)$, from low energies to energies greater than $2
\Delta$. These results predict that the dip feature in the ARPES spectrum in
high temperature superconductors such as BiSrCaCuO, is evidence for the
opening of a spin gap below $T_{c}$.  
\end{abstract}

\newpage 

{\bf Introduction}\\

\vspace{0.25in}

Angle Resolved Photoemission Spectroscopy (ARPES) 
has developed into an important probe of the superconducting
and normal state properties of the cuprate-oxide, high temperature superconductors.
ARPES measurements support LDA predictions for the electronic
bandstructure of the cuprates \cite{Mass} and a $d_{x^2-y^2}$ symmetry
superconducting order parameter \cite{Shen1}. ARPES provides information on how this gap
evolves with doping and its connection to the normal state gap seen in
underdoped cuprates \cite{Shen2}. 

The overall shape of the ARPES spectra
measured in these experiments is not fully understood, however. A common
feature of the data is the co-existence of a broad background and a 
peak. The latter is interpeted as the quasiparticle spectral weight peak
of the Fermi surface electrons emitted from the crystal after the absorption
of the ultraviolet photon ($\simeq 20eV$). The relative
heights of the background and the peak vary from one experiment to another
\cite{Shen1, Onellion, Ding}. In addition to the background, a dip feature
is seen to develop when the crystal is cooled below the superconducting
transition temperature \cite{Shen1}. The connection between this dip feature and a similar
feature observed in tunneling experiments on the cuprates \cite{JZ1,
DCSF, Fischer} has been the subject of recent investigations by the present
authors \cite{Coffey1,Kouznet}. 
The tunneling density of states is another 
measure of the quasiparticle spectral weight. The difference between
tunneling and ARPES arises, in part, because tunneling
measurements provide an average of the quasiparticle spectral weight along a
line of states in the Brillouin Zone, determined by the directional
tunneling matrix element \cite{Kouznet}. ARPES measurements provide
information about the quasiparticle spectral weight in a small
region, determined by experimental resolution,  
around a $\vec k$-space point on the Fermi
surface. 

Recent work \cite{Kouznet} proposed a common explanation for the broad ARPES background
and the linearly increasing tunneling conductances seen in many high temperature
superconductor tunnel junctions \cite{JZ}. It was proposed that the ARPES background is
due to the simultaneous emission or absorption of spin fluctuations by the electron
as it is escapes from the surface layer of the crystal after 
absorbing the ultraviolet photon. The same processes can occur in the
surface layer near the tunnel barrier in a tunnel junction and can lead to a
wide variety of tunneling conductances. Related work has been
carried out by others \cite{China, Kirtley}.

Previous work \cite{Kouznet} made use of a phenomenological model for the spin fluctuation spectral weight
which had been used to fit inelastic neutron scattering
measurements of the spin fluctuation spectrum in YBaCuO \cite{Tranq}. This model
for the spin fluctuation spectrum
did not have superconducting correlations incorporated in it. This issue has 
become relevant due to the observation of structure in the spin fluctuation
spectrum which appears to be directly connected to the development of superconductivity in the cuprates
\cite{Fong}. Another important issue
is the relative height of the main spectral peak and the background as well
as its dependence on the underlying magnitude of the spin fluctuation
susceptibility. The overall magnitude of the inelastic background
contribution to the quasiparticle spectral weight was multiplied by a
constant fitting factor in \cite{Kouznet} in order to reproduce results that compare
favorably with ARPES data.  

In the present work, the spin fluctuation spectral weight,
$D(E)$, which determines the inelastic background, is calculated using the RPA approximation
for the underlying electronic spin susceptibility, $\chi(\vec q, E)$.
The present calculations incorporate the tight binding bandstructure with next nearest
neighbor hopping, appropriate for materials of interest such as BSCCO,
and superconductivity arising from a $d_{x^2-y^2}$ symmetry order parameter.
The effects of the superconducting state on the spin fluctuation spectrum 
are incorporated into a calculation of  
$A(\vec k, E)$ and a comparison is made between predictions and
experimental ARPES measurements of $A(\vec k, E)$
above and below $T_{C}$. 

One of the results of this work is the prediction that the dip feature seen in 
ARPES data below $T_{C}$ on BSCCO \cite{Shen1} is indirect evidence of the development of the 
spin-gap in $\chi(\vec q, E)$ due to the onset of superconductivity. 
The present work allows a controlled investigation to be carried out of
the dependence of the relative magnitude of the inelastic background and the main quasiparticle
peak on the strength of the electron-spin fluctuations. 
Finally, the Van Hove peak
in the underlying tight binding bandstructure  which can strongly influence the energy
dependence of the spin fluctuation spectral weight $D(E)$, is included.

The present model for the quasiparticle spectral weight 
$A(\vec k, E)$ assumes that 
the main peak in the ARPES spectrum arises from
photoelectrons created by the absorption
of ultraviolet photons by electrons at the Fermi energy. The
accompanying broad background is due to the simultaneous
absorption or emission of spin fluctuations by other electrons 
as they
absorb photons.  Spin fluctuation emission dominates at the low temperatures
of interest here.  In the ARPES experimental technique, 
the momentum or $\vec k$ space region being probed 
can be identified with relatively good precision
using energy and momentum conservation.
The photoelectrons which contribute to the inelastic background
are secondary electrons, which having emitted spin fluctuations,
are scattered into the same momentum direction, $\vec k$, 
as those electrons which 
yield the main ARPES spectral peak. They 
are collected by the detector and
labelled with the same momentum $\vec k$ as the main peak yielding an
overall APRES spectrum for a particular $\vec k$ vector.\\

\vspace{0.25in}

{\bf Theoretical Model}\\

\vspace{0.25in}
 
The total quasiparticle spectral weight, $A(\vec k, E)$,  in our model
is made up of two contributions. The main peak is calculated from
\begin{equation}
A^{o}(\vec k, E) = - \frac{1}{\pi} {\rm Im} G(\vec k, E)
\end{equation}
where 
\begin{equation}
G(\vec k,E) = \frac{E + i \Gamma + \xi_{k}}
{(E + i \Gamma)^2 - \xi_{k}^2 - \Delta_{k}^2}
\end{equation}
The electronic bandstructure is defined as 
$\xi_{k}= -2 t (cos(k_{x})+cos(k_{y}))-4 t^{'} cos(k_{x})cos(k_{y}) - \mu$
[Bwhere we have chosen a typical value for the cuprates of $t^{'} =-0.45 t$.
The superconducting order parameter is given by $\Delta_{k}=
\Delta_{o}(cos(k_{x})-cos(k_{y})/2$ with $\Delta_{o}=0.1t$. An electronic damping parameter is
also included through $\Gamma = \Gamma_{o} + \Gamma_{1} (T/T_{C})^3$ with
$\Gamma_{o}$ and $\Gamma_{1}$ chosen to be 0.04t and 0.05t respectively.  
The chemical potential $\mu$ is chosen to be $-1.75t$.

The inelastic contribution is calculated from 
\begin{eqnarray}
A_{inel}(\vec k,E)& =& -\frac{1}{\pi} \int_{-\infty}^{+\infty} dE^{'} Im G(\vec k, E^{'})\\
\nonumber
& & [ D(E-E^{'})(n(E-E^{'})+f(-E^{'})) \Theta (E-E^{'})\\
\nonumber
&+& D(E^{'}-E)(n(E^{'}-E)+f(E^{'}))\Theta(E^{'}-E) ]
\end{eqnarray}
where $n(E)$ denotes the Bose-Einstein distribution and $f(E)$ denotes the Fermi
function. 

The spin fluctuation spectral weight, $D(E)$, is defined as the momentum
integrated spin fluctuation susceptibility multiplied by the square of the 
electron-spin fluctuation coupling constant, $g/t$,
\begin{equation}
D(E) = \frac{1}{(2 \pi)^2} \int_{- \pi}^{+ \pi} dq_{x} 
\int_{-\pi}^{+\pi} dq_{y} (g/t)^{2} Im \chi( \vec q, E)
\end{equation}
where
\begin{equation}
{\rm Im} \chi(\vec q, E) =
\frac{{\rm Im} \chi^{o}(\vec q,E)}{((1-g{\rm Re}\chi^{o}(\vec q,E))^{2} +
(g{\rm Im} \chi^{o}(\vec q,E))^{2})}
\end{equation}
\newpage
The imaginary part of the bare electronic spin susceptibility is defined as
\begin{eqnarray}
{\rm Im} \chi^{o}( \vec q, E) &=& 
\Sigma_{\vec p} \int_{-\infty}^{+\infty}dE^{'}(f(E^{'}+E)-f(E^{'}))\\
\nonumber
& & 
[([u^{2}_{p+q}u^{2}_{p} +u_{p}v_{p}u_{p+q}v_{p+q})]ImG(p+q,E^{'}+E)ImG(p,E^{'}) \\
\nonumber
& +&[u^{2}_{p+q}v^{2}_{p}-u_{p}v_{p}u_{p+q}v_{p+q}]ImG(p+q,E^{'}+E)ImG(p, -E^{'}) \\
\nonumber
& +&[v^{2}_{p+q}u^{2}_{p}-u_{p}v_{p}u_{p+q}v_{p+q}]ImG(p+q, -E^{'} - E)ImG(p,E^{'}) \\
\nonumber
& +&[v^{2}_{p+q}v^{2}_{p}+u_{p}v_{p}u_{p+q}v_{p+q}]ImG(p+q,-E^{'}- E)ImG(p, -E^{'}) ]\\
\nonumber
\end{eqnarray}
where $u_{p}$ and $v_{p}$ represent the usual superconducting coherence
factors and
\begin{equation}
ImG(p,E) = \frac{1}{\pi} \frac{ \Gamma}{(E-E_{p})^2+ \Gamma^{2}}
\end{equation}
where $E_{p}= \sqrt( \xi^{2}_{p} +\Delta^{2}_{p})$.
The real part of $\chi^{o}(\vec q, E)$ is obtained by Kramers-Kronig
transform.

The total quasiparticle spectral weight is then
given by
\begin{equation}
A(\vec k, E) = A^{o}(\vec k, E) + \alpha A_{inel}(\vec k, E)
\end{equation}
where $\alpha$ is an overall constant multiplicative factor which determines 
the relative contributions of the elastic $A^{o}(\vec k, E)$
channel and the inelastic channel $A_{inel}(\vec k, E)$. For the results
presented in this paper, $\alpha$ is chosen to be in the
range from 1 to 4, depending
on the values chosen for the other physical parameters in the calculation
of $A(\vec k, E)$. The choice of $\alpha$ incorporates the relative weighting of
the bulk (leading to $A^{o}(\vec k, E)$) and surface physics (leading to
$A_{inel}(\vec k, E)$).

In our previous \cite{Kouznet} work based on \cite{Duke}, the expression for the total spectral weight
of equations (3) and (8) 
was extracted from the expression for the total elastic
and inelastic tunneling current which can be written as
\begin{equation}
I(V)=  \int  dE N_{S}(E) N_{N}(E-eV)[ f(E) - f(E-eV)]
\end{equation}
where
\begin{equation}
\em{N_{S}(E)}= \Sigma_{\vec k} |T^{el}_{k}|^{2} 
( A^{o}(\vec k, E) + \alpha A_{inel}(\vec k, E))
\end{equation}

In obtaining equations (3) and (8), the tunneling matrix element squared 
in the inelastic contribution to the current, I(V),
(denoted by $|\Lambda^{(1,1)}|^{2}$ in equation (19.37b) of \cite{Duke}) 
was assumed to be given by 
$|\Lambda^{(1,1)}|^{2} = \alpha |\Lambda^{(1,0)}|^{2} (g/t)^{2}$.
By replacing $\epsilon_{L}$ with $\epsilon_{L}-eV$ in equations (19.33b)
and (19.37c) of \cite{Duke}, the combined elastic and inelastic
current can then be written as in equation (9) (using the notation 
$|T^{el}_{k}|^{2}$ in place of $|\Lambda^{(1,0)}|^{2}$). 
The spectral weight 
function of equation (8) yields the density of states
$N_{S}(E)$ using equation (10).

The inelastic tunneling channel
described here implies that an electron tunnels into the superconducting
crystal along a direction determined by the directional tunneling matrix
element $T^{el}_{k}$ and then emits
a spin fluctuation which brings in the spin fluctuation coupling constant
$g/t$ into the overall tunneling matrix element. 
The spectral weight function of equation (8), which 
determines the density of states in equation (10), can then 
be used to interpet ARPES data. 

One approximation in the
present approach is the lack of conservation of momentum.
This approximation is also a feature of related work in this field \cite{China,
Kirtley}. This approximation is valid for the case of sufficiently high
electronic disorder and scattering which will broaden the underyling
$A^{o}(\vec k, E)$. In the clean limit, equation (3) should contain
terms involving
$\Sigma_{\vec q} Im G( \vec k + \vec q, E^{'}) Im \chi(\vec q, E-E^{'})$
within the integral on the right hand side of the equation which would
require a significant increase in numerical computation to yield an 
accurate answer for $A_{inel}(\vec k, E)$. The assumption
inherent in the present work is that $Im G(\vec k + \vec q, E^{'})$
is sufficiently broadened by disorder that it is a reasonable approximation
to take it outside the sum over spin fluctuation wave vectors $\vec q$ and
replace it with $Im G(\vec k, E^{'})$. This approximation can also be
further justified by noting that $Im \chi(\vec q, E-E^{'})$ is strongly
peaked at $\vec Q = ( \pi, \pi)$ and that for the electronic wave vectors
of interest here $Im G(\vec k, E^{'}) \simeq Im G(\vec k + \vec Q, E^{'})$.
We have also tested that the results for the total
quasiparticle spectral weight to be presented 
in the accompanying figures can be generated
with higher values of damping $\Gamma$, and
correspondingly more broadened $A^{o}(\vec k, E)$, than have been
used in the work shown here. 

\vspace{0.25in}

{\bf Results}\\

\vspace{0.25in}

Figures (1) and (2) show results for the spin fluctuation spectral weight $D(E)$ for the
case of a constant coupling constant $g=U$. Figure (1) depicts $D(E)t$ at $T/T_{c}=1.0$ and $0.3$ 
The effect of the onset of superconductivity is evident in the figure in the removal of spectral
weight from below $2 \Delta_{o}$ to higher energies. 

In the figures depicting $D(E)t$ in the present work, the underlying value of 
$\Sigma_{\vec q} Im \chi(\vec q, E)$ is given by dividing $D(E)t$ by $(U/t)^{2}$.
For example, the peak value for $D(E)t$ of approximately
0.3 in figure (1) implies an underlying value of $\Sigma_{q} Im \chi(q, E)$ of 2.0 states per eV assuming
a value of $t=150 meV$. 

In calculating the two curves in figure (1), the sum rule 
\begin{equation}
\frac{1}{2 \pi^{2}} \int d^{2}q \int_{-\infty}^{+\infty} dE 
 \frac{Im \chi(q, E)}{(1 - exp(-E/k_{B}T))} = {\rm constant}
\end{equation}
is imposed. The same value of $U=1.0t$ was used for the two $D(E)$ 
curves in figure (1) with $\mu =-1.75t$.

Figure (2) depicts $D(E)t$ for the case $U=1.5t$ which is close to the largest
possible value of
$U/t$ for the choice 
of $\mu = -1.75t$ before the RPA approximation for $\chi(q,E) $breaks down.  

The resulting quasiparticle spectral weight curves, $A(\vec k, E)t$, 
for these $D(E)t$ curves 
are depicted in figures (3), (4) and (5). All $A(\vec k, E)t$ curves
presented in this paper are calculated for $\vec k$ on the Fermi surface at
$\vec k =(\pi, 0.1624)$. Figure (3) depicts $A(\vec k, E)t$ for 
the $D(E)t$ of figure (1),
with $\alpha =4.0$ in equation (8).
The development of the dip feature associated with the onset of
the spin-gap in $D(E)t$ is clearly visible below the superconducting transition temperature. 

Figures (4) and (5)
depict $A(\vec k, E)t$ using the $D(E)t$ of figure (2) and use
$\alpha =1$ for figure (4) and  
$\alpha=1.5$ for figure (5).  The large increase in magnitude
of $D(E)t$ in figure (2) compared to that of figure (1) allows a sizeable inelastic
background to occur from equation (8) with the elastic and inelastic channels contributing about equally.
Unlike \cite{Kouznet}, the connection between $\alpha$ and the magnitude of
the underlying $Im \chi(q, E)$, which is determined by $U/t$ from equation (5) can now be
explored in a controlled manner. The effect
of interactions in $D(E)t$ is to shift the peak slightly below $2 \Delta_{o}$ as can be seen by comparing
$D(E)$ in figures (1) and (2). This effect can result in a disappearance of the dip feature in
$A(\vec k, E)t$ as can be seen in figure (4). However, a slight increase in $\alpha$ restores the
feature as shown in figure (5). Figure (6) is
generated using the $D(E)t$ of figure (2) with $\alpha=3.0$ and
$\Gamma_{o}=0.06t$ and $\Gamma_{1}=0.075t$ in equation (2). These figures
illustrate the wide variation in the relative heights of the main elastic
peak and inelastic background in $A(\vec k, E)t$ that can be generated within the present
model. A similar variation is seen in ARPES experiments
suggesting that the background is not a bulk phenomenon 
but instead a reflection of the surface physics which can probably vary from sample to sample.

The peak at $E=2 \Delta_{o}$ in $D(E)t$ in figures (1) and (2) is caused by a strong peak in the
underlying $Im \chi(q, E)$ at $\vec q = (\pi, \pi)$ at $E=2 \Delta_{o}$ for $\mu = -1.75t$. This peak
shifts downwards in $E$ as $U$ increases 
due to the part of the denominator involving $Re \chi(q, E)$ in
equation (5) for the susceptibility. As has been just pointed out, the extent to which this occurs can
influence the ability to produce a dip feature in the spectral weight. The peak at $E= 2 \Delta_{o}$ is
also sensitive to the choice of chemical potential $\mu$. A slightly smaller negative value for
$\mu$ will result in the peak moving to higher 
energies in $Im \chi(q, E)$ and diminishing in height \cite{Mazin}. 
The $E= 2 \Delta_{o}$ peak in $Im \chi(q, E)$ is the result of the
underlying Van Hove peak in the tight binding bandstructure and moves to
different values of $E$ as the chemical potential is varied.

Figures (7) and (8) depict $D(E)t$ for the choice $g= -J_{o}( cos(k_{x}) + cos(k_{y}))$ \cite{Levin} 
which enhances the role of the $(\pi, \pi)$ peak in $Im \chi(q, E)$.  
$J_{o}$= 1.0t and 1.2t for figure (7) and $J_{o}$ = 1.2t and 1.4t for figure
(8).
This choice of values for $J_{o}$ ensures that the sum rule of equation (11)
is satisfied \cite{Vilk}.
The resulting $A(k,E)t$ are depicted in figures (9) (which uses the
$D(E)t$ of figure (7)) and (10) (which uses the $D(E)t$ of figure (8)) for 
$\alpha =4.0$ and $\alpha=2.0$ respectively. 
In calculating $A_{inel}(\vec k, E)t$ from equation (3),
the spectral weight $D(E)t$ is integrated over energy $E^{'}$ 
and, as a result, sharp peak structure
in $D(E)t$ is somewhat smeared out in the resulting $A(k,E)t$.

The effect on $A(\vec k, E)t$ of choosing a different point in $\vec k$ space 
has been investigated before in \cite{Kouznet}. Anisotropy as a
function of $\vec k$ enters into the
calculation of the quasiparticle spectral weight in several ways.
Apart from the underlying electronic bandstructure $\xi_{k}$, the
the most significant sources of anisotropy are in the order
parameter $\Delta_{k}$ and in strong coupling effects such as
the quasiparticle damping rate, $\Gamma$. This last issue was treated
phenomenologically in \cite{Kouznet} by increasing the value of
$\Gamma$ for those regions of $\vec k$ space where $\Delta_{k}=0$. The overall
effect is to smear out the main quasiparticle peak and, as a result, 
eliminate the dip feature in the quasiparticle spectral weight.

The spin fluctuation spectral weight $D(E)t$ can also be used to estimate
the quasiparticle damping rate, denoted by $\Gamma$ in equation (2), from
\begin{equation}
\Gamma= 2 \pi \int_{0}^{E_{max}}dE  \frac{D(E)t}{sinh(E/k_{B}T)} 
\end{equation}
Using the $D(E)t$ from figure 1(b) for $T/T_{c}=1$ and assuming
$\Delta_{0}=2 k_{B}T_{C}$, $\Gamma$ is found to be $0.06t$.  This value is comparable to
values of the quasiparticle damping rate $\Gamma$ used in generating
the results of figures (1) to (10).

Equation (12) is derived from the conventional strong coupling
treatment of spin fluctuations. 
The model for the contribution of the inelastic background 
in this paper is different to conventional treatments
of strong coupling effects in calculations of the quasiparticle spectral
weight and tunneling densities of states. The role of spin fluctuations in the
cuprates has been widely investigated in both the normal and superconducting
states\cite{Kampf, MP, MS}. The resulting quasiparticle spectral weights
$A(\vec k, E)$ in these calculations can be used to generate density of states
curves from $N(E)= \Sigma _{k} |T_{k}|^{2}A(\vec k, E)$. In the superconducting state,
densities of states curves will display small corrections relative to the underlying weak coupling densities of
states. This will not provide an explanation for
the rapidly increasing tunneling densities
of states at high bias voltages measured in tunneling experiments on the
cuprates \cite{JZ}. This type of variation with bias voltage is a signature
of an additional inelastic channel, which in the cuprates is assumed to
involve the emission and absorption of spin fluctuations in the surface
region of the sample.

The difference between the inelastic tunneling model and conventional strong
coupling approaches can also be seen by considering the type of
Feynman diagrams that arise in the usual calculation of the tunneling current
using linear response theory. 
Conventional strong coupling corrections
are incorporated with diagrams of the type shown in figure (103) combined
with figure (100a) of \cite{Duke}.  
Inelastic tunneling is calculated using a
different diagram as depicted in figure (98) of \cite{Duke} where both
vertices of the diagram are joined by the propagator representing the
spin fluctuation.\\

\vspace{0.3in}

{\bf Conclusions}\\

\vspace{0.25in}

The spin gap below $E= 2\Delta_{o}$ in $Im \chi(q, E)$ in figures (1), (2),
(7) and (8)
occurs because a spin fluctuation must have an energy $E$ 
greater than or equal to this threshold in order to create 
a quasiparticle-quasihole pair at low temperatures.
The effects of this on the quasiparticle spectral weight and tunneling density of states
have been investigated before in different ways \cite{LiVa,DcLc}. 
\cite{LiVa} incorporated this type of pairbreaking physics phenomenologically in a model
based on the marginal Fermi liquid theory for s-wave superconductivity and investigated the resulting
quasiparticle spectral weight and density of states curves.
In \cite{DcLc}, a quasiparticle damping mechanism based on the same pairbreaking
mechanism 
for a d$_{x^2-y^2}$ order parameter was incorporated approximately into
the quasiparticle spectral weight and the resulting S-I-S current 
characteristics were calculated for
comparison with experiment \cite{DCSF}. 

The work of \cite{Kouznet}, which is the basis for the present
calculations, used a phenomenological model for the spin fluctuation spectral weight. In  
\cite{Kouznet}, a dip feature is present in the quasiparticle 
spectral weight due to a combination of
the narrowing of the main peak, because of the reduction of the scattering rate 
in the superconducting state, and the underlying
shape of the model used for $Im \chi(q,E)$. No superconducting correlations were 
incorporated into the model for $\chi(q,E)$ which are now known to be important \cite{Fong}
and the dip feature was an accidental feature of the model.  

In conclusion, results for a model of the quasiparticle spectral weight 
$A(\vec k, E)t$ have been presented which incorporate a conventional
elastic peak and an inelastic background arising from spin fluctuation
emission processes; equation(3).  The goal is to interpet ARPES measurements
on high temperature superconductors. The results of the present work provide
a model connecting a microscopic calculation of the spin
fluctuation susceptibility $\chi(q, E)$, equations (5) and (6), to the magnitude and
overall shape of the inelastic background seen in ARPES. The spectral
weight curves generated in this approach can also be used to interpet
tunneling conductance measurements on high temperature superconductors
\cite{Kouznet}. The dip feature seen in some ARPES data is caused by 
the development of a spin gap in the underlying spin susceptibility at the
onset of superconductivity.

\newpage

\begin{center}
{\bf Figure Captions}
\end{center}

{\bf All horizontal axes in the figures are in units of E/t.}\\

\vspace{0.25in}

{\bf Figure (1):} The momentum integrated spin fluctuation spectral weight
$D(E)t$ from equation (4) for $g=U=1.0t$.  $T/T_{c}=1.0$ (solid line).
$T/T_{c}=0.3$ (line with symbols). \\

\vspace{0.25in}

{\bf Figure (2):} The momentum integrated spin fluctuation spectral weight
$D(E)t$ from equation (4) for $g=U=1.5t$ for $T/T_{c}=0.3$. \\

\vspace{0.25in}

{\bf Figure (3):} The quasiparticle spectral weight $A(\vec k, E)t$ for
$\vec k$ on the Fermi surface. $\alpha$ of equation (8) is 4.0. This
curve is generated using $D(E)t$ of figure (1). $\vec k$ is chosen to
be $(\pi, 0.1624)$ on the Fermi surface.  $T/T_{c}=1.0$ (solid line).
$T/T_{c}=0.3$ (line with symbols). Curve 3(b) depicts the contributions
of the elastic and inelastic channels separately for $T/T_{c}=0.3$.
The contributions of the two channels depicted in figure 3(b) are typical
of the other $A(\vec k, E)t$ curves in this paper.\\

\vspace{0.25in}                                                           

{\bf Figure (4):} The quasiparticle spectral weight $A(\vec k, E)t$ for
$\vec k$ on the Fermi surface. $\alpha$ of equation (8) is 1.0. This
curve is generated using $D(E)t$ of figure (2). $\vec k$ is chosen to be
$(\pi, 0.1624)$ on the Fermi surface.\\

\vspace{0.25in}                                                         

{\bf Figure (5):} The quasiparticle spectral weight $A(\vec k, E)t$ for
$\vec k$ on the Fermi surface. $\alpha$ of equation (8) is 1.5. This
curve is generated using $D(E)t$ of figure (2). $\vec k$ is chosen to be 
$(\pi, 0.1624)$ on the Fermi surface.\\

\vspace{0.25in}

{\bf Figure 6:} The quasiparticle spectral weight $A(\vec k, E)t$ for
$\vec k$ on the Fermi surface. $\alpha$ of equation (8) is 3.0. This
curve is generated using $D(E)t$ of figure (2). $\Gamma_{o}=0.06t$ and
$\Gamma_{1}=0.075t$ in equation (2). $\vec k$ is chosen to be 
$(\pi, 0.1624)$ on the Fermi surface.\\

\vspace{0.25in}

{\bf Figure (7):} The momentum integrated spin fluctuation spectral weight
$D(E)t$ from equation (4) for $g=J_{q}=-J_{o}(cos(q_{x}+cos(q_{y}))$. 
$T/T_{c}=1.0$, $J_{o}=1.2t$ (dashed line).
$T/T_{c}=0.3$, $J_{o}=1.0t$ (line with symbols).\\

\vspace{0.25in}

{\bf Figure (8):} The momentum integrated spin fluctuation spectral weight
$D(E)t$ from equation (4) for $g=J_{q}=-J_{o}(cos(q_{x}+cos(q_{y}))$. 
$T/T_{c}=1.0$, $J_{o}=1.4t$ (solid line).
$T/T_{c}=0.3$, $J_{o}=1.2t$ (line with symbols).\\

\vspace{0.25in}

{\bf Figure (9):} The quasiparticle spectral weight $A(\vec k, E)t$ for
$\vec k$ on the Fermi surface. $\alpha$ of equation (8) is 4.0. This
curve is generated using $D(E)t$ of figure (7). $\vec k$ is chosen to be
$(\pi, 0.1624)$ on the Fermi surface. 
$T/T_{c}=1.0$ (solid line).
$T/T_{c}=0.3$ (line with symbols).\\

\vspace{0.25in}

{\bf Figure (10):} The quasiparticle spectral weight $A(\vec k, E)t$ for
$\vec k$ on the Fermi surface. $\alpha$ of equation (8) is 2.0. This
curve is generated using $D(E)t$ of figure (8). $\vec k$ is chosen to
be $(\pi, 0.1624)$ on the Fermi surface. 
$T/T_{c}=1.0$ (dashed line).
$T/T_{c}=0.3$ (line with symbols).\\               

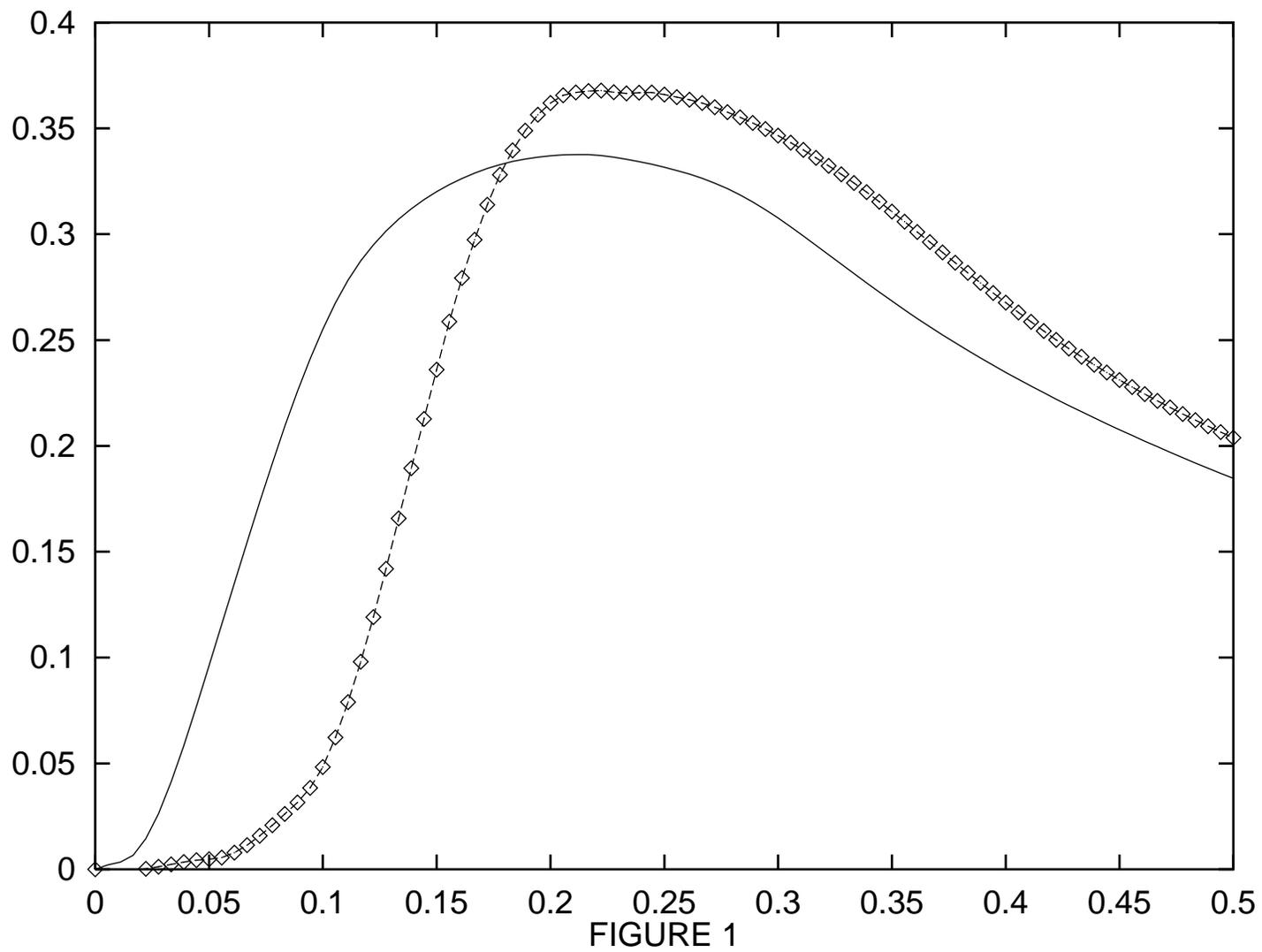
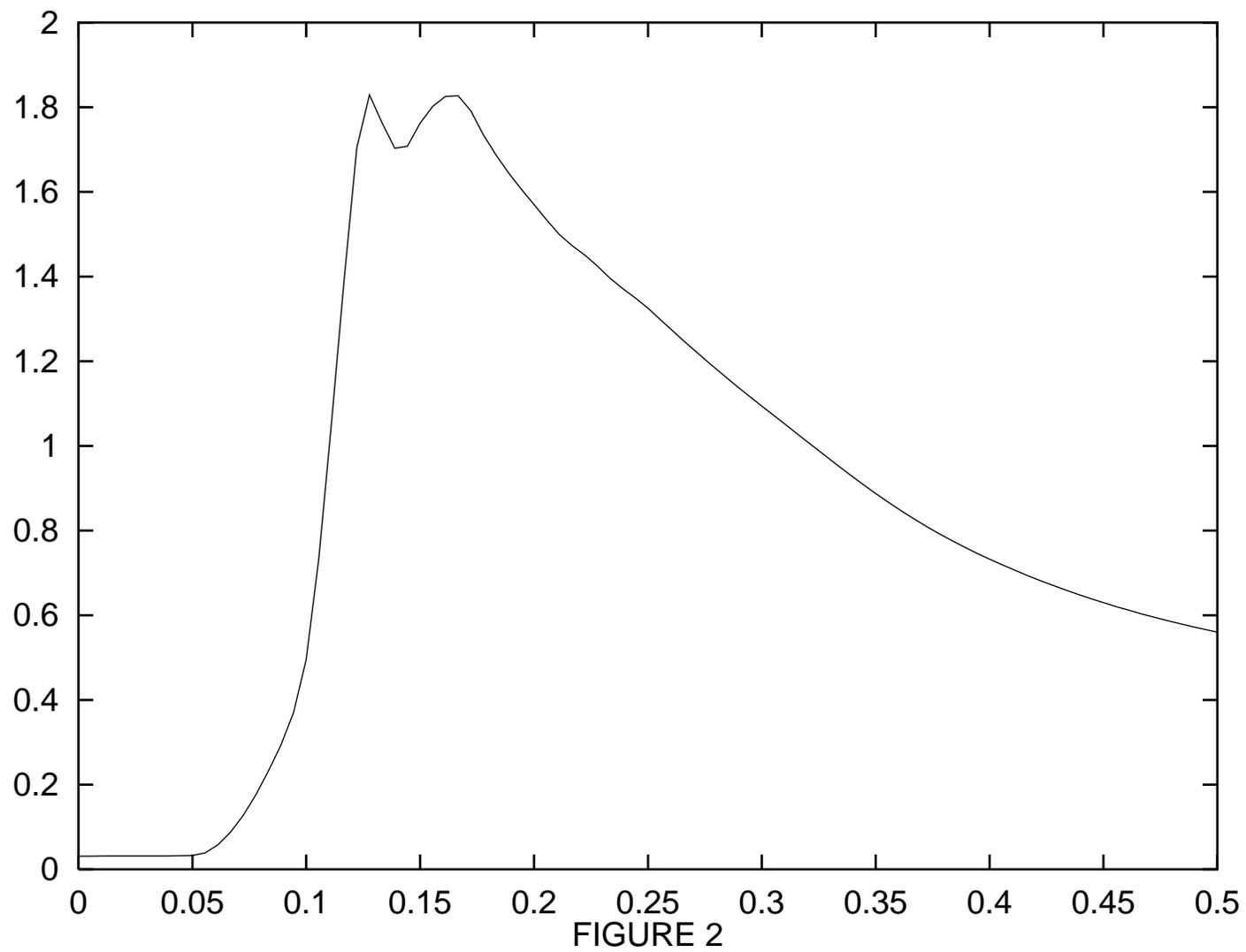
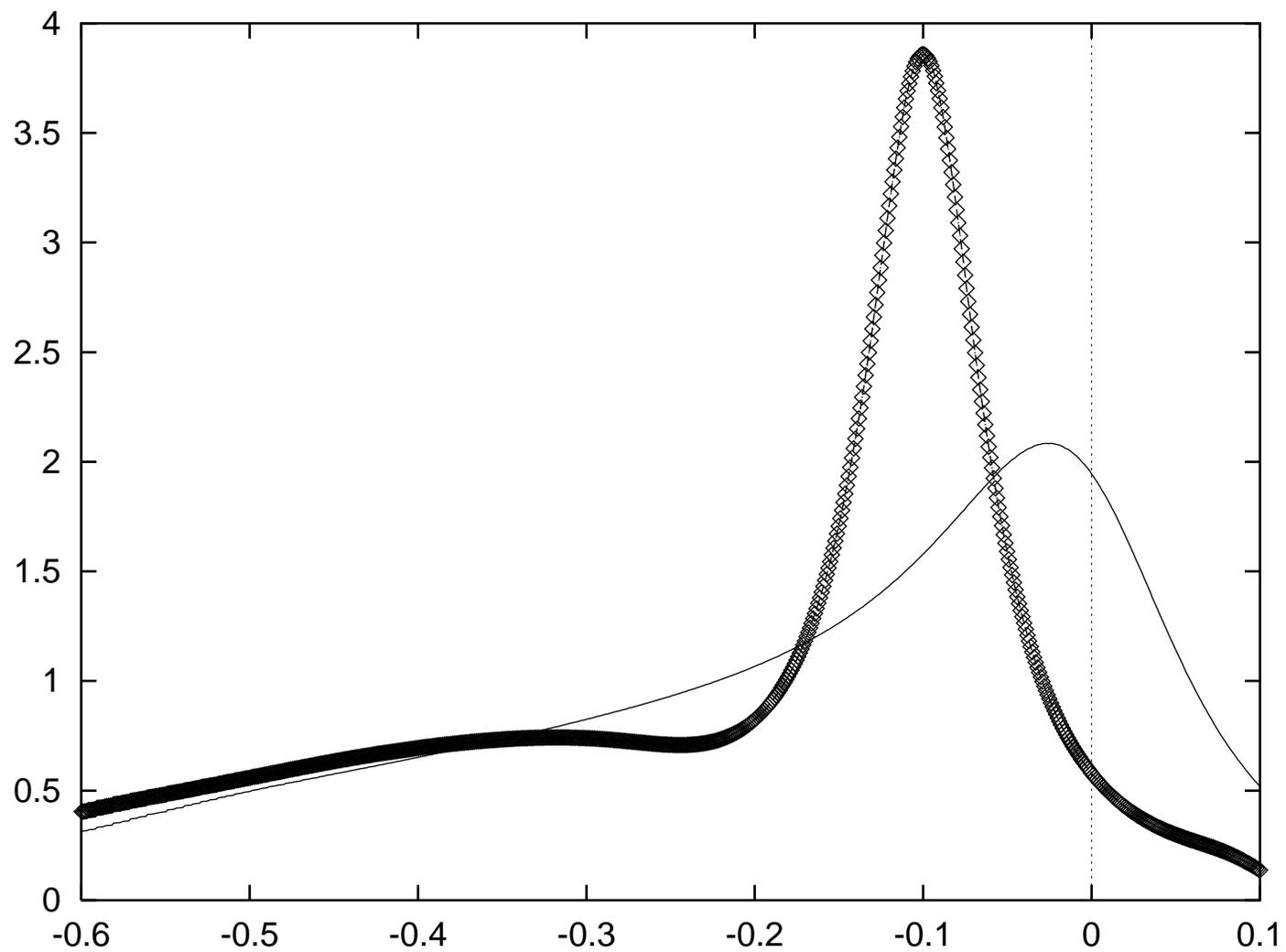
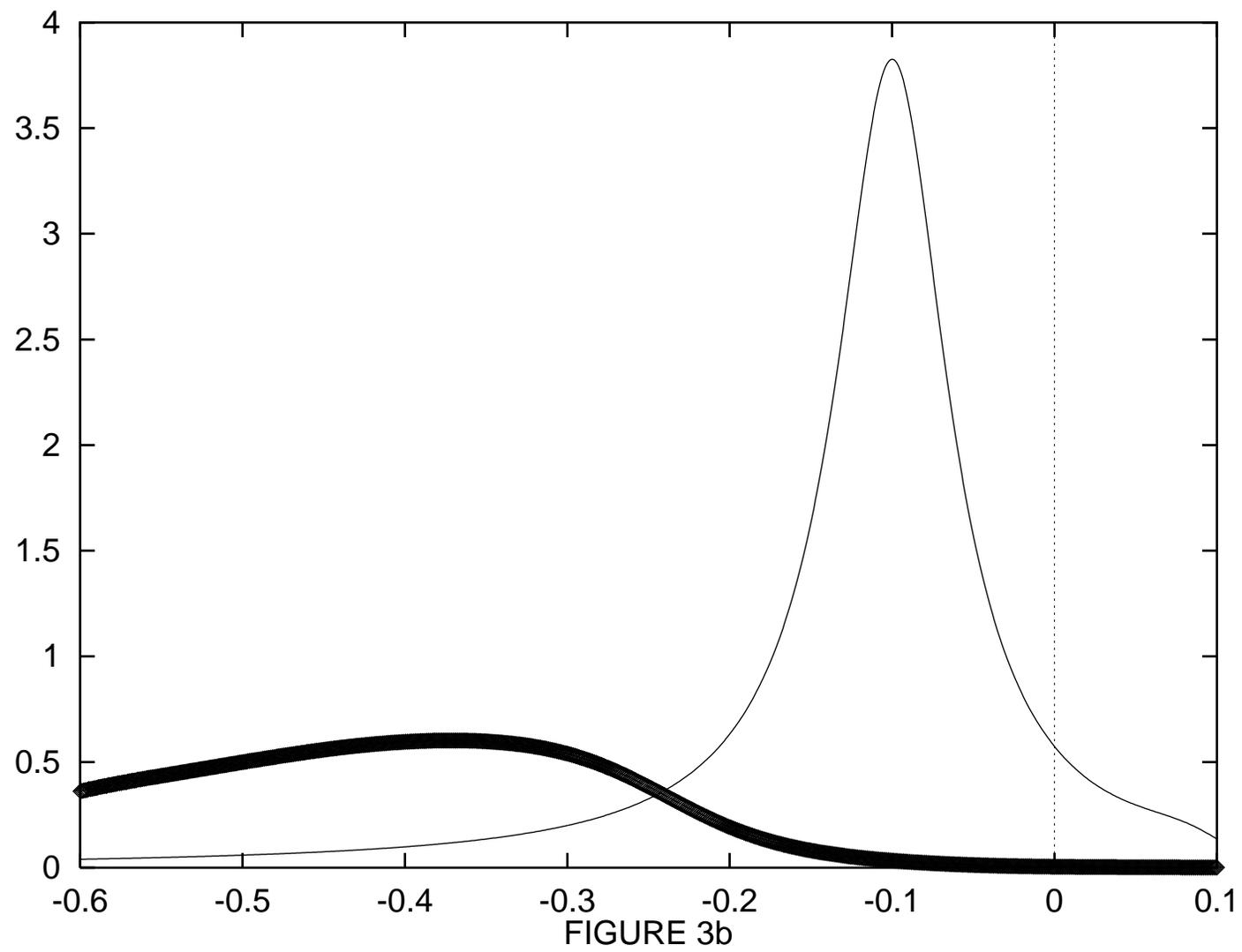
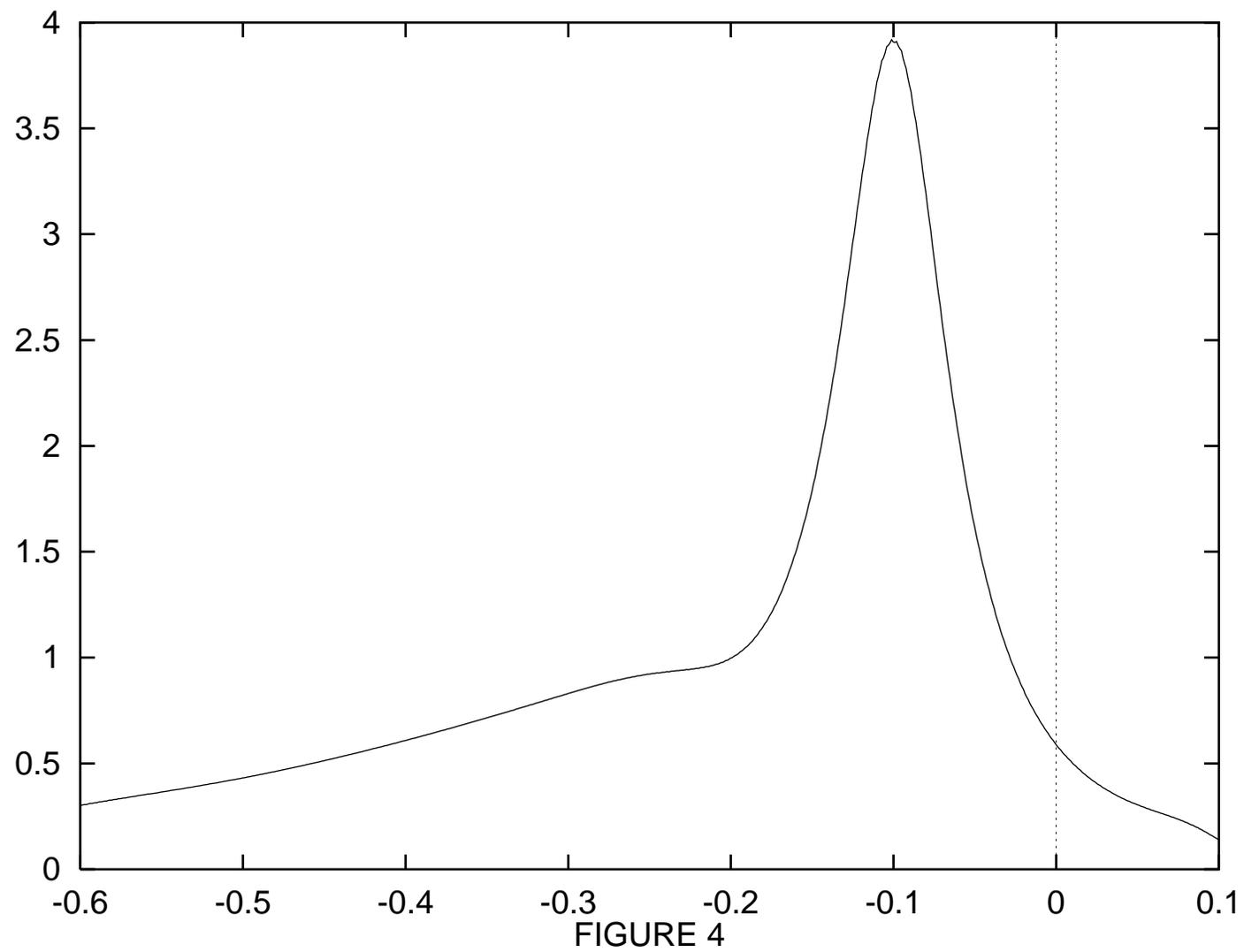
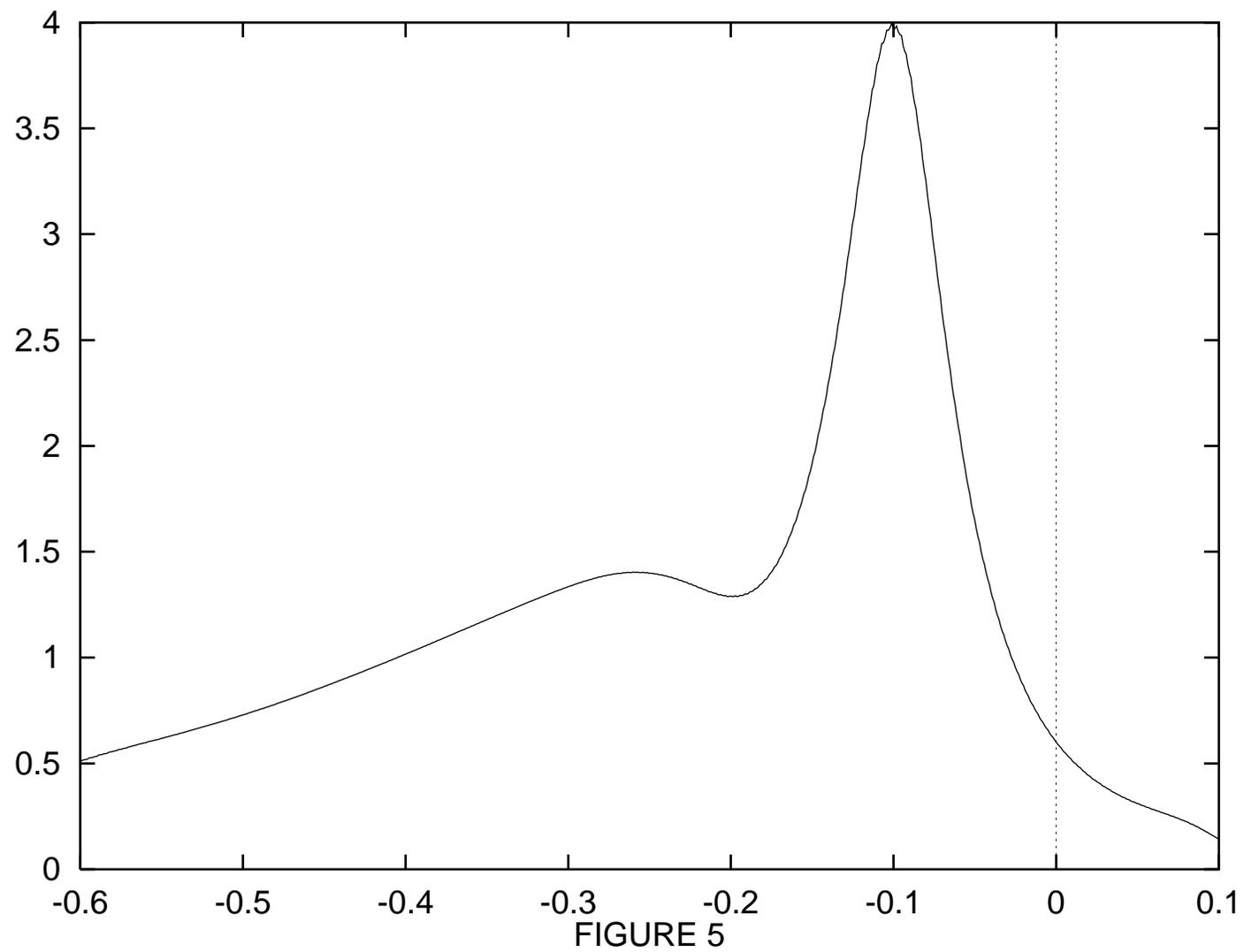
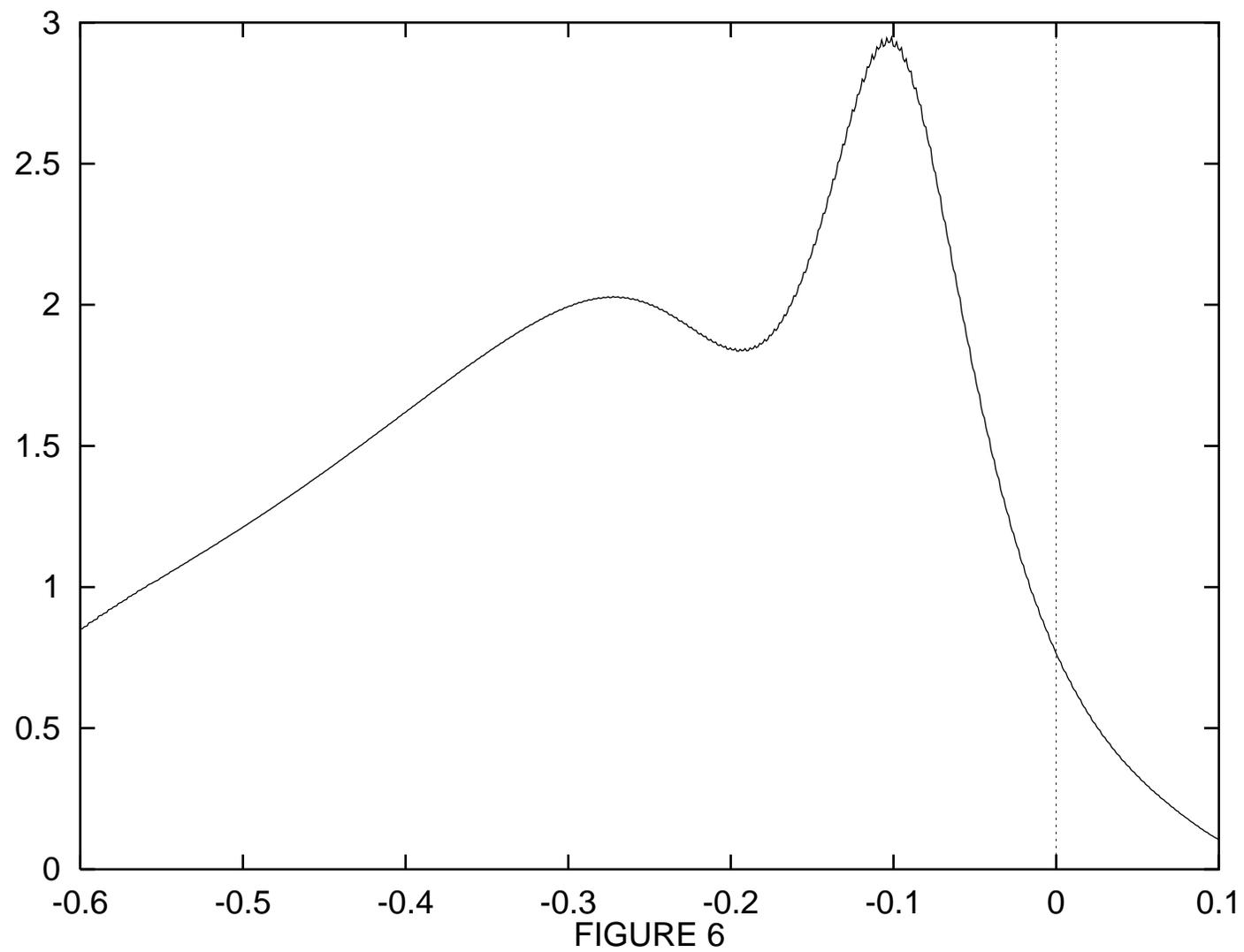
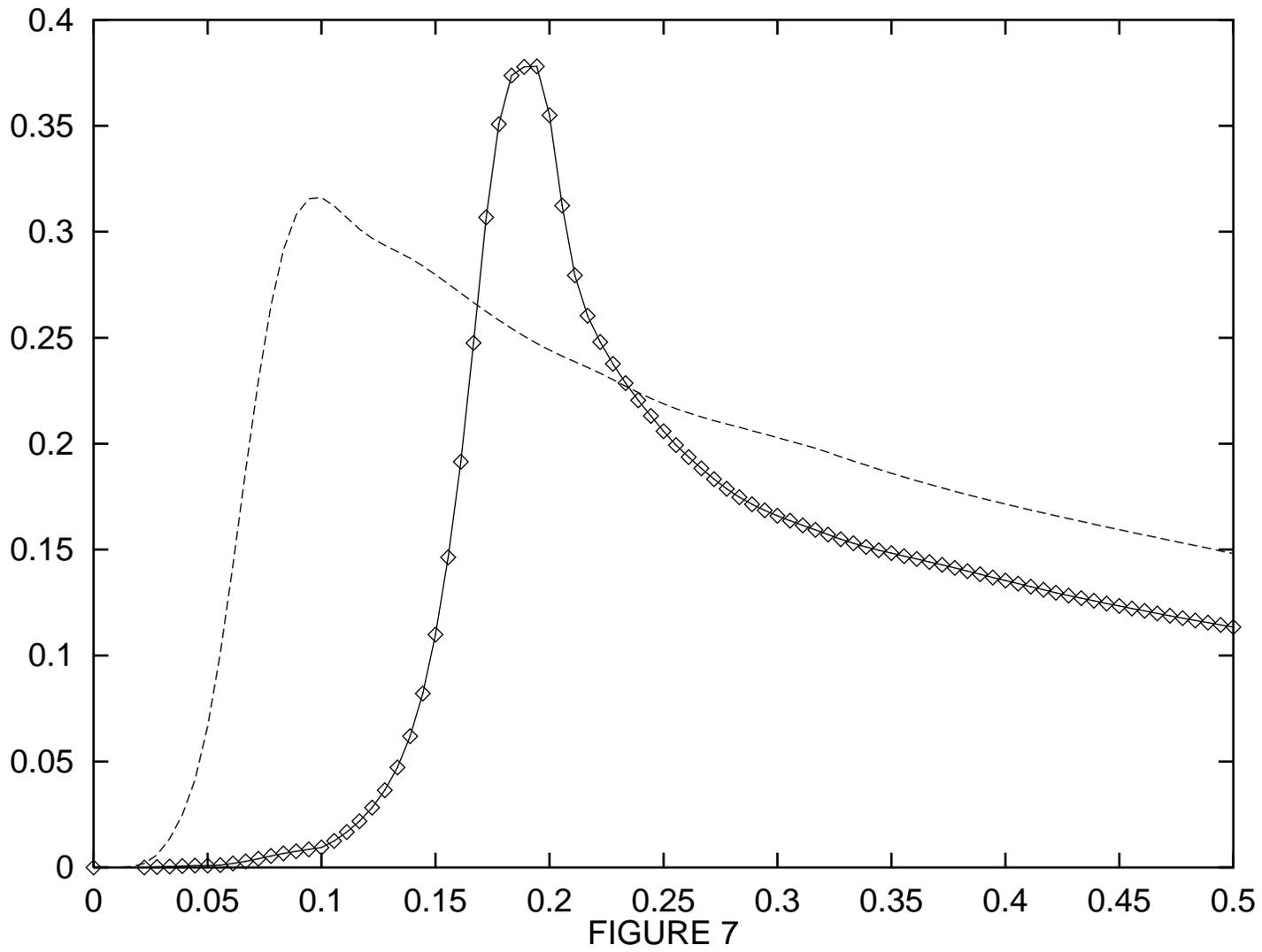
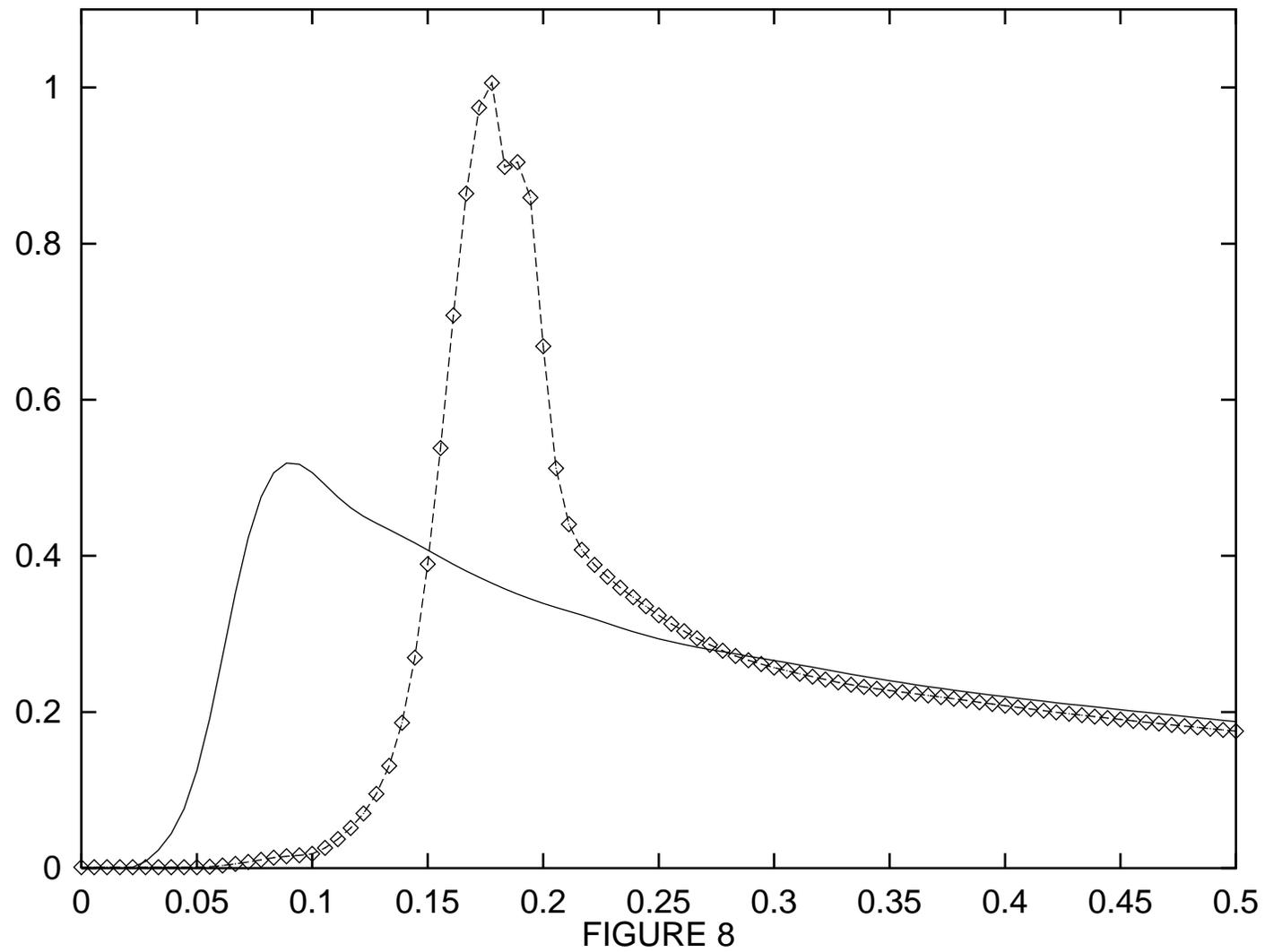
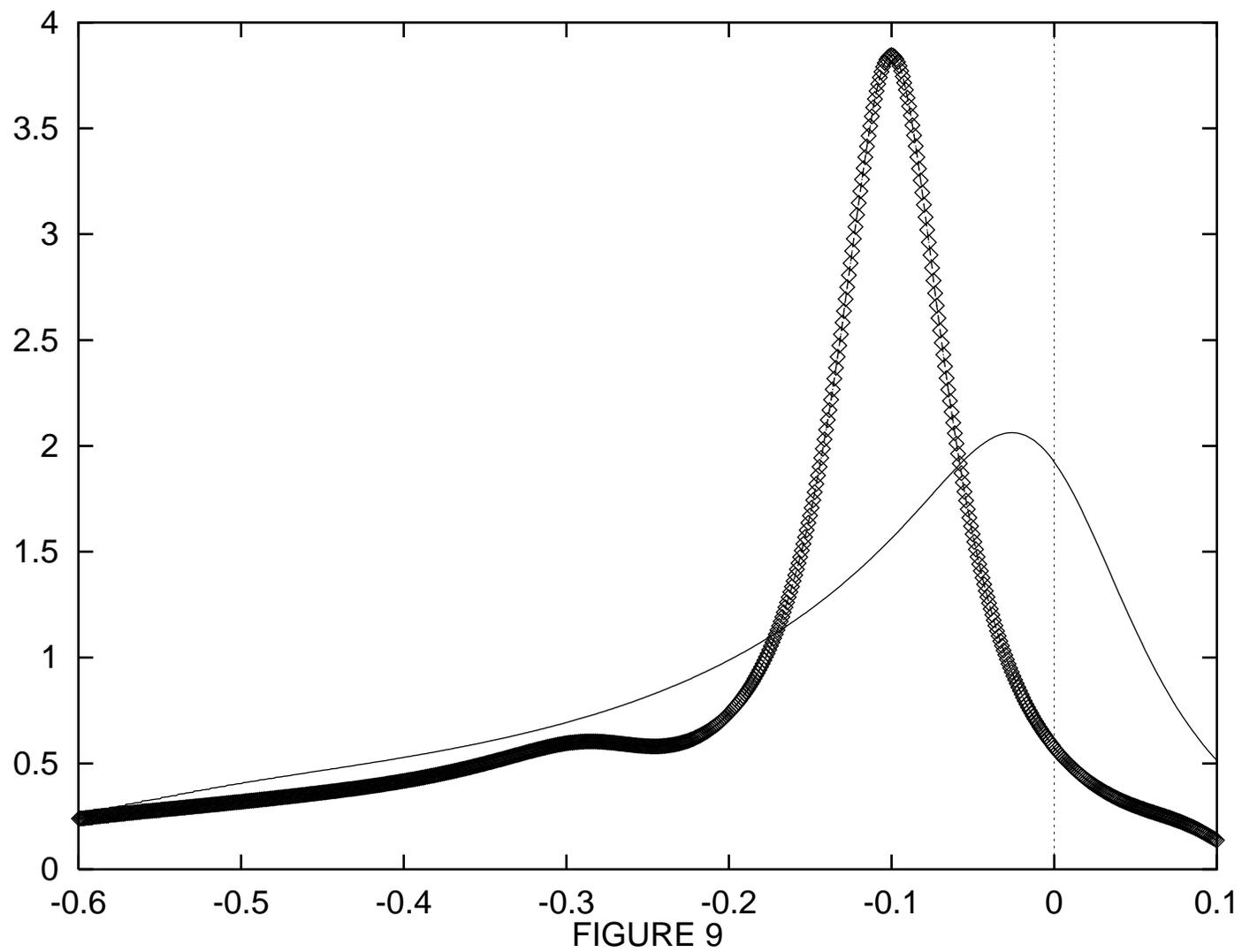
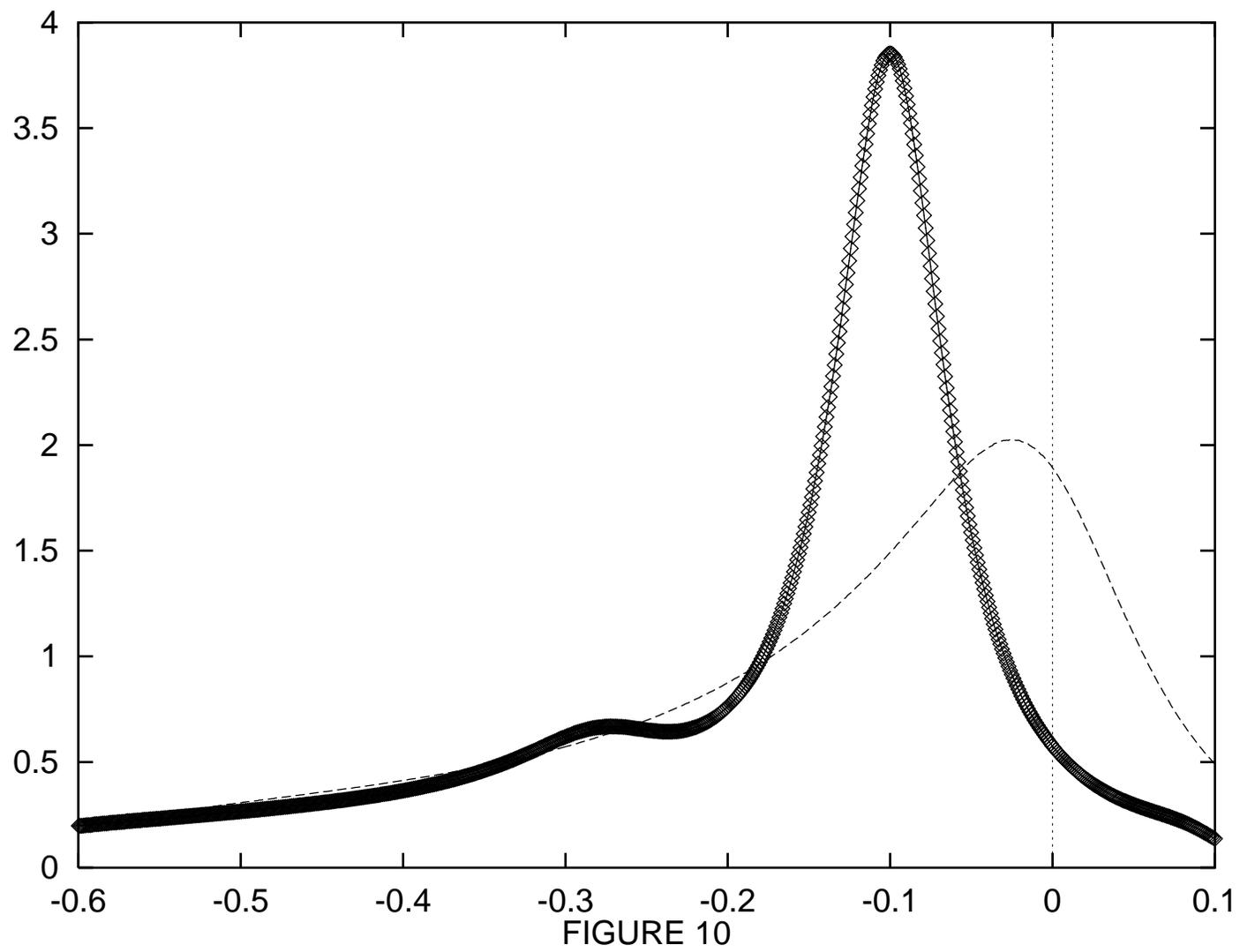
\end{document}